\documentclass[aps,prl,twocolumn,groupedaddress,amsmath,amssymb,showpacs]{revtex4}

\usepackage{graphicx}

\def\e{\mathrm{e}}
\def\i{\mathrm{i}}

\def\a{\alpha}

\def\b{\beta}

\def\e{\epsilon}
\def\g{\gamma}

\def\i{\infty}

\def\ti{\text{i}}

\begin{document}

\title{
Entanglement generation through an open quantum dot: an exact 
approach
}


\author{Takashi Imamura}
\email{timamura@iis.u-tokyo.ac.jp}

\author{Akinori Nishino}
\email{nishino@iis.u-tokyo.ac.jp}

\author{Naomichi Hatano}
\email{hatano@iis.u-tokyo.ac.jp}

\affiliation{Institute of Industrial Science, The University of Tokyo,
4--6--1 Komaba, Meguro-ku, 153--8505, Tokyo}


\date{\today}

\begin{abstract}
We analytically study entanglement generation through an open quantum dot
system described by the two-lead Anderson model.
We exactly obtain the  transition rate
between the non-entangled incident state in one lead and the outgoing spin-singlet 
state in the other lead. 
In the cotunneling process, only the spin-singlet state can transmit.
To discuss such an entanglement property 
in the open quantum system, we construct the exact two-electron scattering state 
of the Anderson model. 
It is striking that the scattering state contains 
spin-singlet bound states induced by the Coulomb interaction. 
The bound state describes the scattering process 
in which the set of momenta is not conserved and  hence it 
is not in the form of a Bethe eigenstate.
\end{abstract}


\maketitle


{\it Introduction:} We present a new exact approach to
electron entanglement generation in an open quantum system.
Entanglement has attracted much attention in
wide range of physics: it is a resource for  
quantum information processing and provides
a new insight into quantum phase transitions 
in statistical physics~\cite{Nielsen-Chuang_05Cambridge,
Amico-Fazio-Osterloh-Vedral_08RMP}. 
In most studies, the entanglement properties
are discussed in closed systems in equilibrium.
In order to study manipulation of 
entanglement, however, we need to consider an open system
out of equilibrium.
Entanglement generation using electrons in  
mesoscopic structures has been 
proposed recently~\cite{Oliver-Yamaguchi-Yamamoto_02PRL,
Saraga-Loss_03PRL,
Lebedev-Blatter-Beenakker-Lesovik_04PRB,
Samuelsson-Sukhorukov-Buttikker_04PRL,
Neder-Ofek-Chung-Heiblum-Mahalu-Umansky_07Nature,
Christ-Cirac-Giedke_08PRB}.
In Refs.~\cite{Oliver-Yamaguchi-Yamamoto_02PRL,
Saraga-Loss_03PRL,Christ-Cirac-Giedke_08PRB}, in particular, 
devices are connected
to reservoirs, electrons enter the device from the reservoirs,
and interactions (the Coulomb interaction as well as the interaction
between electron spin and nuclear spin) are essential for the 
entanglement generation. It is our purpose to discuss  
the entanglement generation in such a situation through 
an exact solution of scattering theory.

\begin{figure}
\includegraphics[width=75mm,clip]{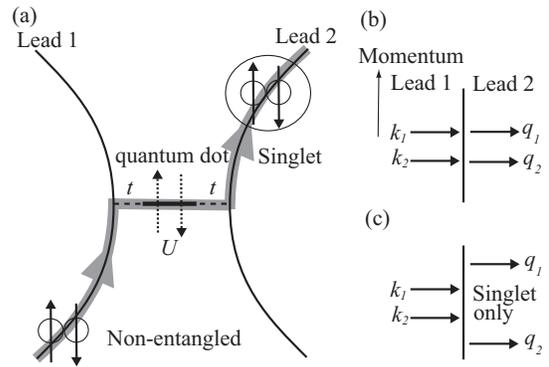}
\caption{\label{fig:anderson} (a) A
schematic diagram of the Anderson model.
(b) A scattering process which conserves the set of momenta.
(c) A cotunneling process.
}
\end{figure}

In this Letter, we obtain an exact result for entanglement property of  
transported electrons of the two-lead  Anderson model. 
The Anderson model is 
a fundamental model describing the electron transport through a 
quantum dot as illustrated in Fig.~\ref{fig:anderson}(a). 
It consists of one energy level 
(the quantum dot) in which two occupying electrons 
with opposite spins interact with each other
(the Hubbard interaction), 
and two leads of noninteracting electrons 
each of which lead is coupled to the dot. 
We calculate the transition rate from the non-entangled incident
state with momenta $k_1$ and $k_2$ on the lead 1 to the singlet 
and triplet states with momenta $q_1$ and $q_2$ on the lead 2.
In the scattering process which conserves the set of momenta 
as in Fig.~\ref{fig:anderson}(b),  
both the triplet and the singlet components of the incident state
can be transmitted to the lead 2.
On the other hand, in the cotunneling process
which conserves the total energy but not the 
set of momenta as in~Fig.~\ref{fig:anderson}(c),
only the singlet component can be transmitted and the triplet
component is filtered out.
We clarify this mechanism by calculating the transition rates
exactly, which is the main achievement of our approach.
The mechanism of the entanglement generation was first
proposed in Ref.~\cite{Oliver-Yamaguchi-Yamamoto_02PRL}; 
the lowest order of our result reproduces their perturbative result. 
An interaction-induced orbital entanglement property 
has been also discussed in a quantum-dot 
system~\cite{Lebedev-Lesovik-Blatter_08PRL}.

For the above purpose, we obtain the exact solution of the  two-electron
scattering state. 
A remarkable point of the state
is that it contains a two-body singlet 
bound state. The bound state is induced 
in the cotunneling process (Fig.~\ref{fig:anderson}(c))
by the Hubbard interaction on the quantum dot. 
A many-body eigenstate of 
the {\it closed} Anderson model can be constructed by the Bethe 
Ansatz~\cite{Andrei_80PRL, Wiegmann_80PLA, Kawakami-Okiji_81PLA,
Kawakami-Okiji_82JPSJ}. In contrast, our exact scattering state 
is a many-body eigenstate of the {\it open} Anderson model
and essentially different from the 
Bethe eigenstate.  
A similar bound state is also discussed 
in Ref.~\cite{Shen-Fan_08preprint}, 
where the two-electron scattering matrix
has been constructed exactly in the Anderson model.
While their study is focused on the asymptotic states of 
electrons which lie far from the quantum dot, our exact 
solution describes electron states both inside
and around the quantum dot.

{\it Model and Result:} The Hamiltonian of the Anderson model is defined as 
$H=H_0+H_1$ where
\begin{align}
H_0&=\sum_{\sigma=\uparrow,\downarrow}\sum_{\ell=1,2}
\int dx
c^{\dagger}_{\ell\sigma}(x)\frac{1}{\ti}\frac{d}
{d x}c_{\ell\sigma}(x)\notag\\
&+\sum_{\sigma=\uparrow,\downarrow}\e_dn_{d\sigma}
+U n_{d\uparrow}n_{d\downarrow},
\label{H0}
\\
H_1&=\frac{t}{\sqrt{2}}
\sum_{\sigma=\uparrow,\downarrow}\sum_{\ell=1,2}
\left( c^{\dagger}_{\ell\sigma}(0)d_{\sigma}+
d^{\dagger}_{\sigma}c_{\ell\sigma}(0)\right).
\label{H1}
\end{align}
Here $c^{\dagger}_{\ell\sigma}(x)~(c_{\ell\sigma}(x))$ denotes the creation
(annihilation) operator of an electron with spin 
$\sigma(=\uparrow,\downarrow)$ on the lead $\ell~(=1,2)$. 
The dispersion relation is linearized to be 
$\e(k)=v_{F}k$  in each lead.
Hereafter we set $v_F=1$. The operator
$d^{\dagger}_{\sigma}~(d_{\sigma})$ represents the creation 
(annihilation) operator on the energy level of a quantum dot 
with energy $\e_d$, and 
$n_{d\sigma}=d^{\dagger}_{\sigma}d_{\sigma}$. The parameter
$t$ represents the coupling between each lead and the dot.
When the energy level is occupied
by two electrons with opposite spins, they feel the 
Coulomb repulsion energy
$U>0$. 

We consider the situation studied 
in Ref.~\cite{Oliver-Yamaguchi-Yamamoto_02PRL}. 
Let $|k_1,k_2;1\rangle$ be the 
non-entangled incident state with momenta $k_1$ and $k_2$ on
the lead 1 defined by
\begin{equation}
\label{inc1}
|k_1,k_2;1\rangle
=c^{\dagger}_{1k_1\uparrow}c^{\dagger}_{1k_2\downarrow}
|0\rangle,
\end{equation}
where $c^{\dagger}_{\ell k\sigma}~(c_{\ell k\sigma})$ denotes the creation (annihilation)
operator of an electron with momentum $k$ and spin $\sigma$ on
the lead $\ell$, and $|0\rangle$ denotes the vacuum state.
We also define a triplet state 
$|q_1,q_2;2,+\rangle$ and a
singlet state $|q_1,q_2;2,-\rangle$ 
with momenta $q_1$ and $q_2$ on the lead 2, 
\begin{equation}
|q_1,q_2,2,\pm\rangle=
\frac{1}{\sqrt{2}}\left(c^{\dagger}_{2q_1\uparrow}
c^{\dagger}_{2q_2\downarrow}\pm c^{\dagger}_{2q_1\downarrow}
c^{\dagger}_{2q_2\uparrow}\right)|0\rangle,
\end{equation}
which are used as outgoing states.
We calculate the transition amplitude between these states:
\begin{equation}
\langle q_1,q_2;2,\pm|T(E_k)
|k_1,k_2;1\rangle\delta(E_k-E_q),
\label{rate}
\end{equation}
where $E_k=k_1+k_2,~E_q=q_1+q_2$ and $T(E)$ represents 
the transition matrix  recursively defined by
\begin{align}
\label{T}
T(E)=H_{1}+H_{1}\frac{1}{E-H_0+\ti 0} T(E).
\end{align}

We obtain new  exact results  for the transition rate:
\begin{align}
\label{tresult}
&\langle q_1,q_2;2,+| T(E_k)
|k_1,k_2;1\rangle\delta(E_k-E_q)\notag\\
&=\frac{ t^2e_{k_1}e_{k_2}}{4\sqrt{2}\ti}
\left(\delta(k_1-q_1)\delta(k_2-q_2)
-\delta(k_1-q_2)\delta(k_2-q_1)\right),
\\
&\langle q_1,q_2;2,-|T(E_k)
|k_1,k_2;1\rangle\delta(E_k-E_q)\notag\\
&=\frac{ t^2e_{k_1}e_{k_2}}{4\sqrt{2}\ti}
\left(\delta(k_1-q_1)\delta(k_2-q_2)
+\delta(k_1-q_2)\delta(k_2-q_1)\right)\notag\\
&+\frac{U}{2\sqrt{2}}\frac{2\e_d-E-\ti t^2}{2\e_d+U-E-\ti t^2}
e_{k_1}e_{k_2}e_{q_1}e_{q_2}\delta(E_k-E_q),
\label{sresult}
\end{align}
where $e_k$ is defined in Eq.~\eqref{ek} below.
Expanding them in the lowest order of $t$, 
we reproduce the perturbative result obtained in
Ref.~\cite{Oliver-Yamaguchi-Yamamoto_02PRL}.

Equation~\eqref{tresult} and the first term in Eq.~\eqref{sresult}
represent the contributions from the scattering process
which conserves the set of momenta $\{k_1,~k_2\}=\{q_1,~q_2\}$
as shown in~Fig.~\ref{fig:anderson}(b).
Note that they vanish if $k_i$ and $q_j$ satisfy the condition of
the cotunneling process (Fig.~\ref{fig:anderson}(c)),
\begin{equation}
E_k=E_q,~~\{k_1,~k_2\}\neq \{q_1,~q_2\}.
\label{cond}
\end{equation}
In contrast, Eq.~\eqref{sresult} has an additional term 
which remains non-zero for $U>0$ under the condition~\eqref{cond}.
The contribution 
appears only in the transition into the singlet state (Eq.~\eqref{sresult}), 
not in the transition into the triplet state (Eq.~\eqref{tresult}).
In other words, we will observe only singlet states if we wait for outgoing electrons
on the lead 2 under the condition~\eqref{cond}.

\begin{figure}
\includegraphics[width=75mm,clip]{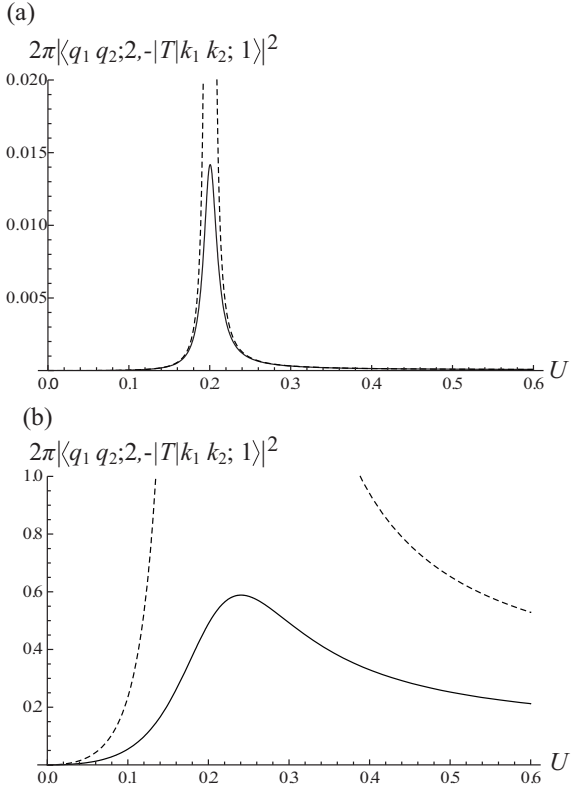}
\caption{\label{fig:transition} 
The transition rate 
$2\pi|\langle q_1,q_2;2,-|T(E_k)|k_1, k_2; 1\rangle|^2$ obtained from
our exact solution (solid line) and a perturbative one (dashed line) for
$\e_d=0$, $k_1=k_2=0.1$, 
$q_1=0.05, q_2=0.15$, $E_k=E_q=0.2$ with (a) $t=0.1$ ($<E_k$) and (b)
$t=0.3$ ($>E_k$).
}
\end{figure}

Figure~\ref{fig:transition} shows the dependence of the transition rate
$2\pi|\langle q_1,q_2;2,-|T(E_k)|k_1, k_2; 1\rangle|^2$ 
on the interaction energy $U$
for $t<E_k$~(Fig.~\ref{fig:transition}(a)) and $t>E_k$
(Fig.~\ref{fig:transition}(b)).
The solid lines indicate our exact result and the dashed lines 
represent the perturbative result in the lowest order of $t$ 
obtained in Ref.~\cite{Oliver-Yamaguchi-Yamamoto_02PRL}.
The perturbative one, being divergent, fails when $U\simeq E_k$
even for $t<E_k$. 

{\it Exact two-electron scattering state:} 
In order to discuss the entanglement generation in such an 
open quantum dot system, what kind of quantum states should 
we focus on?  What is the origin of the second term in Eq.~\eqref{sresult}?

We find that the two-electron 
scattering state $|\phi\rangle$
given by the solution of the Lippmann-Schwinger equation,
\begin{equation}
|\phi\rangle=|k_1,k_2;1\rangle
+\frac{1}{E_k-H_0+\ti 0}H_1|\phi\rangle
\label{L}
\end{equation}
plays an important role. 
In fact we have the relation
 \begin{equation}
T(E_k)|k_1,k_2;1\rangle =H_1|\phi\rangle,
\label{Tphi}
\end{equation}
which enables us to calculate the transition amplitude~\eqref{rate}.

In the remaining part of this paper,  we construct the exact 
solution of $|\phi\rangle$.
In Ref.~\cite{Nishino-Imamura-Hatano_09PRL}, 
we obtain the exact solution of many-electron scattering scattering states
in a spinless model. Here we apply the same technique to the two-lead 
Anderson model. The importance of $|\phi\rangle$ results from the fact that 
the system is an {\it open} quantum system. It is suitable for the
study of transport properties in  the open quantum system, but the exact solution
of such a state has not been discussed very much. As we will explain later,
our exact solution has a remarkable property that 
the  Bethe eigenstate does not have.

By applying the transformation
$c_{e\sigma}(x)=({c_{1\sigma}(x)
+c_{2\sigma}(x)})/{\sqrt{2}},~
c_{o\sigma}(x)=({c_{1\sigma}(x)
-c_{2\sigma}(x)})/{\sqrt{2}},
$
we decompose the Hamiltonian into $H=H_e+H_o$, where
\begin{align}
H_e&= \sum_{\sigma=\uparrow,\downarrow}
\left(
\int 
dx 
c^{\dagger}_{e\sigma}(x)\frac{1}{\ti}\frac{d}
{d x}c_{e\sigma}(x)
+\e_dn_{d\sigma}\right)
+U n_{d\uparrow}n_{d\downarrow}
\notag\\
&
+ \sum_{\sigma=\uparrow,\downarrow}t \left( c^{\dagger}_{e\sigma}(0)d_{\sigma}+
d^{\dagger}_{\sigma}c_{e\sigma}(0)\right), 
\label{ehamiltonian}
\\
H_o&=\sum_{\sigma=\uparrow,\downarrow}
\int dx
c^{\dagger}_{o\sigma}(x)\frac{1}{\ti}\frac{d}{d x}c_{o\sigma}(x).
\label{ohamiltonian}
\end{align}
Note that the odd part $H_o$ is completely decoupled from the
even part $H_e$. Thus the two-lead Anderson model~\eqref{H0} and~\eqref{H1} 
can be transformed to the one-lead Anderson model~\eqref{ehamiltonian} 
with the free part~\eqref{ohamiltonian}.  
Let $|k_1,k_2; s \rangle_{\a\b}$, $(\a,\b)=(e,e),(e,o),(o,o)$, $s=+,-$
denote the scattering eigenstate with the incident state being
a two-electron plain-wave state 
\begin{equation}
|k_1,k_2; \pm\rangle_{\a\b}^{\text{(i)}}=\frac{1}{\sqrt{2}}
(c^{\dagger}_{\a k_1\uparrow}c^{\dagger}_{\b k_2\downarrow}
\pm c^{\dagger}_{\a k_1\downarrow}c^{\dagger}_{\b k_2\uparrow})
|0\rangle.
\label{inc}
\end{equation}
The scattering state $|\phi\rangle$
is expressed as~\cite{Nishino-Imamura-Hatano_09PRL}
\begin{align}
|\phi\rangle &=\frac{1}{2\sqrt{2}}\Big(
\sum_{s=\pm}\big(
|k_1,k_2;s\rangle_{ee}+|k_1,k_2;s\rangle_{eo}+|k_1,k_2;s\rangle_{oo}\big)
\notag\\ 
&-|k_2,k_1;+\rangle_{eo}+|k_2,k_1;-\rangle_{eo}\Big),
\label{phieo}
\end{align}
because substituting the incident states~\eqref{inc} for all states on the
right-hand side gives the incident state~\eqref{inc1}.

Before giving the two-electron scattering eigenstate, 
we first mention the one-electron eigenstate
$|k\sigma\rangle_{e/o}$ defined as a solution 
of the Schr{\"o}dinger equation
$H_{e/o}|k\sigma\rangle_{e/o}=k|k\sigma\rangle_{e/o}$. 
The one-electron eigenstate can 
be expressed as follows:
\begin{align}
&|k\sigma\rangle_{e}=\left(\int dx g_{k}(x)
c^{\dagger}_{e\sigma}(x)+ e_k 
d^{\dagger}_{\sigma} \right)|0\rangle,\\
&|k\sigma\rangle_o = \int dx h_{k}(x)
c^{\dagger}_{o\sigma}(x)|0\rangle,
\end{align} 
where the eigenfunctions are given by
\begin{align}
&g_k(x)=
\frac{1}{\sqrt{2\pi}}e^{\ti kx}\left(\theta(-x)+\theta(x)
\frac{e_k}{e^*_k}\right),
\label{gk}\\
&e_k=\frac{1}{\sqrt{2\pi}}\frac{t}{k-\e_d+\ti t^2/2},
\label{ek}
\\
&h_{k}(x)=\frac{1}{\sqrt{2\pi}}e^{\ti kx}.
\label{gkodd}
\end{align}
Here $\theta(x)$ is the step function and $e_k/e^*_k$
in Eq.~\eqref{gk} represents the phase factor due to  
the scattering by the dot.

Next we consider the two-electron scattering state, which is
written in the form
\begin{align}
&|k_1,k_2;s=\pm \rangle_{\a\b}\notag\\
&=\Bigg(\int dx_1dx_2 g_{\a\b}^{s}(x_1,x_2;k_1,k_2)
c^{\dagger}_{\a\uparrow}(x_1)c^{\dagger}_{\b\downarrow}(x_2)
\notag\\
&+\int dx e_{\a\b}^{s}(x;k_1,k_2)
\left(c^{\dagger}_{\g\uparrow}(x)
d^{\dagger}_{\downarrow}\pm 
c^{\dagger}_{\g\downarrow}(x)d^{\dagger}_{\uparrow}
\right)\notag\\
&+f_{\a\b}^{s}(k_1,k_2)
d^{\dagger}_{\uparrow}d^{\dagger}_{\downarrow}\Bigg)
|0\rangle.
\label{twoparticle}
\end{align}
Here $\g=e$ for $(\a,\b)=(e,e)$, and $\g=o$ 
otherwise.
From Eqs.~\eqref{Tphi} and~\eqref{phieo},  we find that 
the transition rate~\eqref{rate} is expressed in terms of the eigenfunctions:
\begin{align}
&\langle q_1,q_2;2,\pm|T(E_k)
|k_1,k_2;1\rangle\delta(E_k-E_q)\notag\\
&=\delta(E_k-E_q)\frac{t}{8\sqrt{2\pi}}\int_{-\i}^\i dx
(h_{q_1}(x)\mp h_{q_2}(x))
\notag\\
&\times(e_{ee}^{\pm}(x;k_1,k_2)-e_{eo}^{\pm}(x;k_1,k_2)\pm e_{eo}^{\pm}
(x;k_2,k_1)).
\label{tratee}
\end{align}
We construct the eigenfunctions specifically $e_{\a\b}^{s}(x)$ 
for all cases of $(\a,\b)$ to calculate the transition rate~\eqref{tratee}.
First of all, let us
consider the case $\a=\b=e,~s=-$. 
Only in this case, the eigenfunctions depend on the Coulomb interaction
 $U$. The functions with total energy $E_k=k_1+k_2$ are obtained by solving 
the two-electron
Sch{\"o}dinger equation
\begin{align}
\label{Sch}
&\left(\frac{1}{\ti}\left(\partial_1
+\partial_2\right)-E_k\right)g_{ee}^{-}(x_1,x_2)\notag\\
&+t\left(\delta(x_1)e_{ee}^{-}(x_2)+\delta(x_2)e_{ee}^{-}(x_1)
\right)=0,\\
&\left(\frac{1}{\ti}\frac{d}{d x}
+\e_d-E_k \right)e_{ee}^{-}(x)
+tg_{ee}^{-}(x,0)+t\delta (x) f_{ee}^-=0,\\
&(2\e_d+U-E_k)f_{ee}^-+2te_{ee}^{-}(0)=0
\label{Sch2}
\end{align}
under the condition 
\begin{equation}
g_{ee}^{-}(x_1,x_2)
=\frac{1}{2\sqrt{2}\pi}\left(e^{\ti (k_1x_1+k_2x_2)}+e^{\ti (k_1x_2+k_2x_1)}
\right)
\label{ourgform}
\end{equation}
for~$x_1, x_2<0$.
The result is as follows:
\begin{align}
&g_{ee}^{-}(x_1,x_2)=\sum_{Q}\Big(\frac{1}{\sqrt{2}}g_{k_1}(x_{Q_1})g_{k_2}(x_{Q_2})\notag\\
&+\frac{\sqrt{2}t^2Ue_{k_1}e_{k_2}}
{2\e_d+U-E_k-\ti t^2}e^{\ti E_kx_{Q_2}}Z(x_{Q_1Q_2})\theta (x_{Q_1}) 
\Big),
\label{gxx}
\\
&e_{ee}^{-}(x)=\frac{1}{\sqrt{2}}
\left(e_{k_1}g_{k_2}(x)+e_{k_2}g_{k_1}(x)\right)\notag\\
&\hspace{9mm}+\frac{\sqrt{2}\ti tUe_{k_1}e_{k_2}}
{2\e_d+U-E_k-\ti t^2}e^{\ti E_kx}Z(x),
\label{ex}
\\
&f_{ee}^{-}=\sqrt{2}e_{k_1}e_{k_2}-
\frac{\sqrt{2}Ue_{k_1}e_{k_2}}{2\e_d+U-E-\ti t^2},
\label{f}
\end{align}
where $Q=(Q_1, Q_2)$ is a permutation of (1,2),  
$x_{ij}=x_{i}-x_{j}$ and
\begin{align}
&Z(x)=e^{(\ti \e_d+t^2/2) x}\theta(-x).
\label{bound}
\end{align}
It is obvious that Eq.~\eqref{gxx} satisfies 
the condition~\eqref{ourgform}.
Note that the condition~\eqref{ourgform} corresponds to 
the plane-wave incident state~\eqref{inc}. We 
confirm that our solution~\eqref{gxx}--\eqref{f} of the 
Schr{\"o}dinger equation~\eqref{Sch}--\eqref{Sch2}
 satisfies
the Lippmann-Schwinger equation,
\begin{equation}
|k_1,k_2;-\rangle_{ee}=|k_1,k_2;-\rangle_{ee}^{\text{(i)}}
+\frac{1}{E_k-H_0+\ti 0}H_1|k_1,k_2;-\rangle_{ee}.
\label{Lk}
\end{equation}

Remarkable is that Eqs.~\eqref{gxx} and~\eqref{ex} contain 
the {\it two-body bound state} $Z(x)$ in Eq.~\eqref{bound}. 
The range of binding is $t^{-2}$,  which itself is independent
of $U$. 
Note the following properties: (i) It describes a scattering process that does not conserve the momentum set  as the cotunneling process~\eqref{cond}. 
In contrast, the first term of Eq.~\eqref{gxx} represents the direct ($k_1=q_1, k_2=q_2$) and exchange $(k_1=q_2,k_2=q_1)$ processes.
(ii) It is induced by the Coulomb interaction $U$;
It vanishes for $U=0$.
(iii) It only appears in the case $\a=\b=e$ and $s=-$
since $f_{\a\b}^{s}$ in Eq.~\eqref{twoparticle} vanishes 
in the other cases
and thereby the eigenfunctions 
do not depend on  $U$.
From these properties, we find that it is this bound state that produces 
the second term of Eq.~\eqref{sresult}. 

This solution is essentially different from the Bethe eigenstate~\cite{Andrei_80PRL, Wiegmann_80PLA, Kawakami-Okiji_81PLA,
Kawakami-Okiji_82JPSJ}.  
In the Bethe ansatz, 
we would suppose that $g_{ee}^-(x_1,x_2)$ is in the 
following form:
\begin{equation}
g_{ee}^-(x_1,x_2)=\sum_QA_Qg_{k_1}(x_{Q_1})g_{k_2}(x_{Q_2}).
\label{gform}
\end{equation}
Here $A_Q$ depends on $k_i$ and the parameters of the model 
($\e_d, t$ and $U$). Note that the solution is characterized by
the fixed set $\{k_1,k_2\}$.
Although $k_i$ could be a complex number,
it is obvious that Eq.~\eqref{gform} 
cannot take the form of Eq.~\eqref{gxx}.

In the different cases of $\a,\b,s$ from $\a=\b=e,s=-$,
the scattering states are interaction-free eigenstates.
In particular, $e^s_{\a\b}(x)$  are obtained as follows:
\begin{align}
&e_{ee}^{+}(x)=\frac{1}{\sqrt{2}}
\left(-e_{k_1}g_{k_2}(x)+e_{k_2}g_{k_1}(x)\right),
\label{+exf}
\\
&e_{eo}^{\pm}(x)=\mp\frac{1}{\sqrt{2}}e_{k_1}h_{k_2}(x),
~e_{oo}^{\pm}(x)=0.
\label{+ooexf}
\end{align}
The scattering 
state $|\phi\rangle$ constructed from Eq.~\eqref{phieo} with
these eigenfunctions is shown to 
satisfy the Lippmann-Schwinger equation~\eqref{L}.
Using all these exact solutions in Eq.~\eqref{tratee}, we finally 
arrive at the desired results~\eqref{tresult} and~\eqref{sresult}.

{\it Conclusion:}
In this Letter we have constructed the exact two-electron scattering 
state
and discussed  its entanglement 
property. For the exact calculation of
the quantities~\eqref{tresult} and~\eqref{sresult}, 
our solution  is
essential. 
We have clarified that the electron transport through
the quantum dot has a potential advantage for entanglement
generation. 
The construction of many-electron scattering states in other
more complicated models and the understanding of their 
entanglement property are interesting issues in the future. 

Our exact scattering state
could also be a powerful tool to understand non-equilibrium electron transport with
finite bias voltage.
In a spinless model, the analytic approach for this topic has 
been proposed~\cite{Mehta-Andrei_06PRL, Nishino-Hatano_07JPSJ,
Doyon_07PRL, Boulat-Saleur_08PRB} and the importance 
of  many-electron scattering states
on the nonequilibrium current has been pointed 
out  recently~\cite{Dhar-Sen-Roy_08PRL,Roy-Soori-Sen-Dhar_09preprint,
Nishino-Imamura-Hatano_09PRL}.
In particular, our approach in 
Ref.~\cite{Nishino-Imamura-Hatano_09PRL} succeeded in 
obtaining nonperturbative result on nonlinear current-voltage 
characteristics. This suggests that our exact many-electron scattering state 
yields important information  about the Kondo effect out of equilibrium~\cite{
Hershfield-Davies-Wilkins_91PRL,
GoldhaberGordon_98Nature,
Cronenwett_98Science,
Wiel-Franceschi-Fujisawa-Elzerman-Tarucha-Kouwenhoven_00Science}.

The authors would like to thank Prof.~T.~Fujii for discussions.
One of the authors (T.I.) also would like to thank Prof.~Y.~Yamamoto
for helpful comments.
The present study is partially supported by 
Grant-in-Aid for Young Scientists (B) No.~20740217,
Grant-in-Aid for Scientific Research (B) No.~17340115,
and CREST
of Japan Science and Technology Agency.


\end{document}